\documentclass{article}
\usepackage{amsmath}

\input{tcilatex}

\begin{document}

\title{Extended BRST symmetries. Quantum approach}
\author{Radu Constantinescu, Carmen Ionescu \\
Department of Theoretical Physics \and University of Craiova, Romania}
\maketitle

\begin{abstract}
The aim of this lecture is to present in a comprehensible way what the BRST
quantization means and how the "classical" master equation, action and BRST
transformations have to be prolonged towards the same "quantum" items. The
presentation will focus not only on the standard BRST symmetry, but on
larger symmetries as $sp(2)$, both in the Lagrangean and in the Hamiltonian
formalisms. How to find answers to these questions in more sophisticated
cases will be illustrated by the example of a nonlinear system with open
superalgebra.
\end{abstract}

\section{Introduction}

All fundamental interactions from nature are described by theories with
internal symmetries or gauge theories. Such theories, as QED, QCD,
electro-weak theory and gravity, ask for a special approach because of the
unphysical degrees of freedom involved by the gauge invariance. For example,
when quantizing such theories the direct computation of the path-integrals
is meaningless since the integration over gauge directions in the measure
would make the path-integral infinite-valued. The BRST technique \cite{1} is
one of the approaches which allows to overcome this dificulty. It assumes
the replacement of the local (gauge) symmetry with a global (BRST) one. The
BRST\ symmetry is expressed either as a differential operator $s$, or in a
canonical form, by the antibracket $(~,~)$ in the Lagrangean
(Batalin-Vilkovisky)\ case \cite{BV}\ and by the extended Poisson bracket $%
\{~,~\}$ in the Hamiltonian (Batalin-Fradkin-Vilkovisky)\ formulation \cite%
{BFV}:%
\begin{equation}
s\ast =(\ast ,S)=\{\ast ,\Omega \}.  \label{a8}
\end{equation}%
The BRST generator $S$ and the BRST\ charge $\Omega $ are both defined in
extended spaces generated by the real and by the ghost-type variables. To
perform path-integral calculations in this frame, it is necessary to remove
the redundant gauge variables, that is to gauge-fix the action by choosing a
gauge fermion $Y.$Elimination of gauge variables assures the BRST invariance
of the action \cite{HenTeit} but not of the measure. A BRST transformation
of the coordinates could generate non-trivial terms in the action. These
terms can be exponentiated and adsorbed in the action (\ Fadeev-Popov
trick). One obtain an extended action called "quantum" action.

In this paper we will explicitly show how the BRST formalism allows to
construct well defined path-integrals and what is the concrete form of the
quantum master equation in the special case of a physical system described
by a nonlinear gauge algebra. The interest on such systems was initiated by
the discovery of conformal field theories \cite{3} which led to a new class
of gauge theories with nonlinear gauge algebras, the so-called $\mathcal{W}%
_{N}$ algebras \cite{4}. The BRST approach for non-linear superalgebras was
developed in \cite{Lav}.

In the standard BRST approach the gauge fixing procedure asks for a
"non-minimal sector" which suposes the introduction of new, supplementary
variables. In the extended symmetries this non-minimal sector appears in a
natural way. Now, the gauge fixing function can be constructed in a more
simple way and the gauge fixing action is obtained easily. Singura
difficulty which can be appear in the case of the extended BRST formalism is
asigurarea of the invariance of the integrating measure, so that the average
of the observables to be independent de the choice of the gauge. Ne propunem
in this paper to show how can done this in the case of gauge theories with
nonlinear gauge superalgebras.

\section{Quantum Master Equations }

\subsection{The Standard BRST\ Approach}

There are two main procedures in obtaining a quantum description of the
gauge systems: \textit{the Dirac quantization method} and the \textit{path
integral method}. In the Hamiltonian formalism the existence of the gauge
symmetries is equivalent with the existence of some constraints imposed to
the theory and this is why the Dirac quantization method, which is
especially apropriate for the description of systems with constraints, is a
very useful method in quantizing gauge theories. As we intend to built up
the quantum equation for interacting fields, we will concentrate our
approach on the path integrals' formalism which is more suitable for the
quantum description of these systems. This\ approach to the quantization of
systems with symmetries, both in the Lagrangean (BV) formalism and in the
Hamiltonian (BFV) one, started from the Fadeev-Popov trick and ended with
the development of the BRST\ technique. The gauge fixing procedure assures
the BRST invariance of the gauge fixed action but not automatically of the
generating functional $Z_{Y}$ and of the expectation values $\left\langle
F\right\rangle _{Y}$. For a gauge theory with an action $S[\phi ,\phi ^{\ast
}]$ we have:%
\begin{equation}
Z_{Y}=\int d\phi ^{A}\exp (\frac{i}{\hbar }S_{\Psi }[\phi ^{A},\phi
_{A}^{\ast }=\frac{\delta Y}{\delta \phi ^{A}}])  \label{a9}
\end{equation}%
and%
\begin{equation}
\left\langle F\right\rangle _{Y}=\int d\phi ^{A}F\exp (\frac{i}{\hbar }%
S_{Y}).  \label{a10}
\end{equation}%
The\ independence on the choice of $Y$ and the BRST invariance of (\ref{a9})
and (\ref{a10}), essential requirements for a consistent quantum
description, asks more for the invariance of the measure $d\phi ^{A}$ to the
BRST transformation:%
\begin{equation}
\phi ^{A}\rightarrow \phi ^{\prime A}=\phi ^{A}+(-1)^{\varepsilon _{A}}(\phi
^{A},S)\mu =\phi ^{A}+(-1)^{\varepsilon _{A}}\frac{\delta S}{\delta \phi
_{A}^{\ast }}\mu  \label{a11}
\end{equation}%
where $\mu $ is a constant, anti-commuting parameter. In the general case
the transformation (\ref{a11}) leads to 
\begin{equation}
d\phi ^{\prime A}=(1-(-1)^{\varepsilon _{A}}\frac{\delta }{\delta \phi ^{B}}%
\frac{\delta S}{\delta \phi _{B}^{\ast }}\mu )d\phi ^{A}=(1-(\Delta S)\mu
)d\phi ^{A}  \label{a121}
\end{equation}%
and nontrivial terms in the measure could be generated. The invariance
condition would impose:%
\begin{equation}
d\phi ^{\prime A}=d\phi ^{A}\Rightarrow \Delta S=0;\Delta \equiv
(-1)^{\varepsilon _{A}}\frac{\delta }{\delta \phi ^{A}}\frac{\delta }{\delta
\phi _{A}^{\ast }}.  \label{a12}
\end{equation}%
The condition (\ref{a12}) might not be fulfilled. Although, all the terms
generated by $(\Delta S)\mu )d\phi ^{A}$ can be exponentiated and absorbed
as "quantum corrections" of the action $S$:%
\begin{equation}
W=S+\hbar W_{1}+\hbar ^{2}W_{2}+\cdots  \label{a13}
\end{equation}%
One obtains the "quantum action" $W$ which satisfies a Quantum Master
Equation (QME), an extension of the classical master equation $(S,S)=0$. It
has the form:%
\begin{equation}
\frac{1}{2}(W,W)-i\hbar \Delta W=0.  \label{a14}
\end{equation}%
The equation (\ref{a14}) can be written in the equivalent form%
\begin{equation}
\Delta e^{\frac{i}{\hbar }W}=0.  \label{a15}
\end{equation}%
It is important to note that $\Delta $ is not a true derivation, its action
beeing given by the following rule: 
\begin{equation}
\Delta (\alpha \beta )=\alpha \Delta \beta -(-)^{\varepsilon \alpha }(\Delta
\alpha )\beta -i\hbar \Delta (\alpha ,\beta ).  \label{a151}
\end{equation}%
The classical BRST operator $s$ can be, it also, extended to a quantum one: 
\begin{equation}
s\ast \equiv (\ast ,S)\rightarrow \sigma \ast \equiv (\ast ,W)-i\hbar \ast
\label{a16}
\end{equation}%
The quantum version of the BRST invariance will be expressed by: 
\begin{equation}
\sigma F=0\Longleftrightarrow \Delta (Fe^{\frac{i}{\hbar }W})=0.  \label{a17}
\end{equation}

\subsection{The $sp(2)$ extended BRST Quantization}

Because of some difficulties the standard BRST\ approach met in the gauge
fixing procedure, procedure which suposes the introduction of some
supplementary variables from a \textit{non-minimal sector}, \textit{extended
BRST symmetries} has been formulated. The main such extension is known as
the $sp(2)$\textit{\ BRST symmetry} and suposes the existence of two
anticommuting differential operators, $s_{1}$ and $s_{2},$ which can be
joined in a symplectic doublet $s_{1}$,$s_{2}$ with: 
\begin{equation}
s=s_{1}+s_{2};~s^{2}=0.  \label{02}
\end{equation}

In the Hamiltonian formalism \cite{cls7}, the $sp(2)$ BRST symmetry is
canonical generated by the BRST charges $\Omega _{a},\ a=1,2$ with $%
\varepsilon (\Omega _{a})=1$. By defining an extended phase-space and  a
generalized Poisson bracket, the BRST charges will be given by:%
\begin{equation}
s_{a}\ast =\{\ast ,\Omega _{a}\};a=1,2.  \label{021}
\end{equation}%
Their concrete form depends on the theory and it can be obtained by using
the homological perturbations theory \cite{HenTeit}. The extended
Hamiltonian $H$ is the solution of the equations%
\begin{equation*}
\{H,\Omega ^{a}\}=0,\ a=1,2
\end{equation*}%
with boundary condition%
\begin{equation*}
\left. H\right\vert _{Q=\lambda =0}=H_{0}(q,p).
\end{equation*}%
It is obtained using the same homological perturbations theory \cite{HenTeit}%
. The gauge fixing procedure leads to a gauge fixed Hamiltonian of the form%
\begin{equation*}
H_{Y}=H+\frac{1}{2}\varepsilon _{ab}\{\Omega ^{a},\{\Omega ^{b},Y\}\}
\end{equation*}%
where $Y$ is the gauge fixing functional defined in terms of the real and
ghost type coordinates.

Even if it is easier to develop the Hamiltonian formalism, it does not
always lead to a covariant gauge fixing action. This is way the passage to
the Lagrangean formalism is indicated. It is in the Lagrangean frame where
we should study the invariance \ of the generating functional $Z_{Y}$ and of
the expectation values of the form (\ref{a10}) for observables.

In the Lagrangean case \cite{cls7'}, two antibrackets are defined and the
master equations take the form:

\begin{equation}
\frac{1}{2}(S,S)_{a}+V_{a}S=0;~a=1,2  \label{03}
\end{equation}%
where $\{V_{a},a=1,2\}$ represent two special "non-canonical" operators.
There is a direct modality of obtaining the $sp(2)$ BRST Lagrangean
formalism, but it assumes the use of a very large spectrum of ghost
generators. To avoid this unuseful extension, it is simpler at the classical
level to construct the Lagrangean formalism following its equivalence with
the Hamiltonian one \cite{noi}. We will present how the classical $sp(2)$
theory can be extended at quantum level in the frame of path integrals
formalism. We will not give here more details but we direct to \cite{5},
where quantum $sp(2)$ master equations were obtained in the Hamiltonian and
then in the Lagrangean schemes. In this quantum context, the invariance
condition (\ref{a12}) has as correspondent the equations:

\begin{equation}
\Delta ^{a}S=0;\Delta ^{a}\equiv (-1)^{\varepsilon _{A}}\frac{\partial ^{r}}{%
\partial \phi ^{A}}\frac{\partial ^{r}}{\partial \phi _{Aa}^{\ast }}%
+V^{a}=(-1)^{\varepsilon _{A}}\frac{\partial ^{r}}{\partial \phi ^{A}}\frac{%
\partial ^{r}}{\partial \phi _{Aa}^{\ast }}+\varepsilon ^{ab}\phi
_{Ab}^{\ast }\frac{\partial ^{r}}{\partial \overline{\phi }_{A}}  \label{04}
\end{equation}%
and%
\begin{equation}
\Delta ^{a}\Delta ^{b}+\Delta ^{b}\Delta ^{a}=0,\ a,b=1,2.  \label{s40}
\end{equation}%
If $\Delta ^{a}S\neq 0$, the operators $\Delta ^{a},a=1,2$ determine the $%
sp(2)$- QME in the $sp(2)$ symmetric formulation of gauge theories:%
\begin{equation}
\Delta ^{a}e^{\frac{i}{\hbar }W}=0\Leftrightarrow \frac{1}{2}%
(W,W)^{a}=i\hbar \Delta ^{a}W,\ \ a=1,2  \label{06}
\end{equation}%
where $W$ represents the "quantum action", the $\hbar -$order corrections
being involved by integrating measure. In the next section we will come back
to this problem, approaching it in more concrete and applied manner.

\section{The Example of Nonlinear Superalgebras}

Let us consider a system described in a phase-space $M\equiv
\{q^{i},p_{i},i=1,\cdots ,n\}$ with $\varepsilon (q^{i})=\varepsilon
(p_{i})=\varepsilon _{i}$ by a set of first class constraints $\{G_{\alpha
},~\alpha =1,\cdots ,m\}$ with $\varepsilon (G_{\alpha })=\varepsilon
_{\alpha }$ and by the canonical action 
\begin{equation}
S_{can}[q,p,u]=\int dt[\overset{.}{q}^{i}p_{i}-H^{(0)}(q,p,u)],\ i=1,\cdots
,n  \label{2.351}
\end{equation}%
\begin{equation}
H^{(0)}(q,p,u)=H_{0}(q,p)+u^{\alpha }G_{\alpha }.  \label{2.352}
\end{equation}%
The Lagrange multipliers $\{u^{\alpha },\alpha =1,\cdots ,m\}$ will play the
key rol in the establishment of the equivalence between the extended BRST
Hamiltonian and Lagrangean formalisms. We suppose that the constraints
satisfy involution relations of the form:%
\begin{equation}
\{G_{\alpha },G_{\beta }\}=G_{\gamma }f_{\alpha \beta }^{\gamma }+G_{\gamma
}G_{\delta }g_{\alpha \beta }^{\gamma \delta }  \label{d0}
\end{equation}%
The gauge algebra is given by the relations (\ref{d0}) and by%
\begin{equation}
\{H_{0},G_{\alpha }\}=G_{\beta }V_{\alpha }^{\beta }+G_{\delta }G_{\beta
}U_{\alpha }^{\beta \delta }.\   \label{d1}
\end{equation}%
We consider the case when $f_{\alpha \beta }^{\gamma }$, $g_{\alpha \beta
}^{\gamma \delta }$, $V_{\alpha }^{\beta }$, $U_{\alpha }^{\beta \delta }$
are true constants. We deal in this case with a quadratic non-linear Lie
algebra, case which correspond to a constrained system with open algebra 
\cite{7}. An example of such system is the Yang-Mills fields theory with the
nilpotent BRST charge a quadratic function of the Fadeev-Popov ghost fields.

We will develop the $sp(2)$ BRST Hamiltonian formalism for the previous
nonlinear superalgebra. The extended phase space will be generated in this
case by the real variables $\{q^{i},p_{i},i=1,...,n\}$ and by the ghost-type
variables $\{Q^{\alpha a},\mathcal{P}_{\beta b},\lambda ^{\alpha },\pi
_{\beta }\}$. The last ones satisfythe relations:%
\begin{equation*}
\{Q^{\alpha a},\mathcal{P}_{\beta b}\}=\delta _{\beta }^{\alpha }\delta
_{b}^{a},\ \{\lambda ^{\alpha },\pi _{\beta }\}=\delta _{\beta }^{\alpha
}\delta _{b}^{a}
\end{equation*}
On the basis of the Jacobi identities for the structure functions of the
superalgebra \cite{Lav} we can show that for any superalgebras (\ref{d0}), (%
\ref{d1}):%
\begin{equation*}
\Omega ^{a}=G_{\alpha }Q^{\alpha a}+\varepsilon ^{ab}P_{\alpha b}\lambda
^{\alpha }+\frac{1}{2}(-)^{\varepsilon _{\alpha }}\mathcal{P}_{\gamma
c}(f_{\alpha \beta }^{\gamma }+G_{\delta }g_{\alpha \beta }^{\delta \gamma
})Q^{\beta b}Q^{\alpha a}+
\end{equation*}%
\begin{equation*}
+\frac{1}{2}\pi _{\tau }\lambda ^{\sigma }Q^{\gamma a}(f_{\sigma \gamma
}^{\tau }+G_{\delta }g_{\sigma \gamma }^{\delta \tau })(-)^{\varepsilon
_{\alpha }+\varepsilon _{\sigma }+1}(-)^{\varepsilon _{\sigma }\varepsilon
_{\gamma }}+\frac{1}{8}[(f_{\alpha \beta }^{\gamma }+G_{\delta }g_{\alpha
\beta }^{\delta \gamma })(f_{\sigma \rho }^{\alpha }+G_{\delta }g_{\sigma
\rho }^{\delta \alpha })\cdot 
\end{equation*}%
\begin{equation*}
\cdot (-)^{(\varepsilon _{\beta }+1)(\varepsilon _{\alpha }+\varepsilon
_{\rho })+\varepsilon _{\beta }\varepsilon _{\sigma }}+(f_{\sigma \alpha
}^{\gamma }+G_{\delta }g_{\sigma \alpha }^{\delta \gamma })(f_{\beta \rho
}^{\alpha }+G_{\delta }g_{\beta \rho }^{\delta \alpha })(-)^{(\varepsilon
_{\sigma }+1)(\varepsilon _{\rho }+\varepsilon _{\alpha })+\varepsilon
_{\alpha }\varepsilon _{\rho }}+
\end{equation*}%
\begin{equation*}
+(f_{\rho \beta }^{\alpha }+G_{\delta }g_{\rho \beta }^{\delta \alpha
})(f_{\sigma \alpha }^{\gamma }+G_{\delta }g_{\sigma \alpha }^{\delta \gamma
})(-)^{(\varepsilon _{\sigma }+1)(\varepsilon _{\rho }+\varepsilon _{\alpha
})+\varepsilon _{\alpha }\varepsilon _{\rho }}+
\end{equation*}%
\begin{equation}
+(f_{\sigma \beta }^{\alpha }+G_{\delta }g_{\sigma \beta }^{\delta \alpha
})(f_{\alpha \rho }^{\gamma }+G_{\delta }g_{\alpha \rho }^{\delta \gamma
})(-)^{(\varepsilon _{\rho }+1)(\varepsilon _{\alpha }+\varepsilon _{\beta
})+\varepsilon _{\rho }\varepsilon _{\sigma }}]\varepsilon _{bc}\pi _{\gamma
}Q^{\beta a}Q^{\rho b}Q^{\sigma c}.  \label{d11}
\end{equation}

If in (\ref{2.351}) we consider the momenta $\{p^{i},i=1,\cdots ,n\}$ as
auxiliarly variables, we can eliminate them on the basis of their equations
of motion. One obtain the action 
\begin{equation}
S_{0}[q,u]=\int dt\;L_{0}(q,\overset{.}{q},u)  \label{2.354}
\end{equation}%
where the Lagrange multipliers $u^{\alpha },\ \alpha =1,\cdots ,m$ are seen
now as real fields. The action (\ref{2.354}) is invariant at the gauge
transformations%
\begin{equation}
\delta q^{i}=a_{\alpha }^{i}(q,\overset{.}{q})\varepsilon ^{\alpha },\
\delta u^{\alpha }=\overset{.}{\varepsilon }^{\alpha }-(V_{\beta }^{\alpha
}+G_{\delta }U_{\beta }^{\delta \alpha })\varepsilon ^{\beta }+(f_{\alpha
\beta }^{\gamma }+G_{\delta }g_{\alpha \beta }^{\gamma \delta })u^{\gamma
}\varepsilon ^{\beta }.  \label{2.355}
\end{equation}%
The Noether's identities will have the form 
\begin{equation*}
\frac{\delta S_{0}}{\delta q^{i}}a_{\beta }^{i}+\frac{\delta S_{0}}{\delta
u^{\alpha }}[-(V_{\beta }^{\alpha }+T_{\delta }U_{\beta }^{\delta \alpha
})+(f_{\beta \gamma }^{\alpha }+G_{\delta }g_{\beta \gamma }^{\delta \alpha
})u^{\gamma }]-\frac{d}{dt}(\frac{\delta S_{0}}{\delta u^{\beta }})=0.
\end{equation*}%
Starting from (\ref{2.354}) we will develop the $sp(2)$ BRST Lagrangean
formalism \cite{cls7'}. The complete spectrum of the antifields is given in
our case by%
\begin{equation}
Q_{\mathcal{A}a}^{\ast }\equiv \{Q_{Aa}^{\ast },u_{\alpha a}^{\ast
}\}=\{q_{ia}^{\ast },Q_{\alpha ab}^{\ast },\lambda _{\alpha a}^{\ast
},,u_{\alpha a}^{\ast },\ a,b=1,2\},  \label{s1}
\end{equation}%
\begin{equation}
\overline{Q}_{\mathcal{A}}\equiv \{\overline{Q}_{A},\overline{u}_{\alpha
}\}=\{\overline{q}_{i},\ \overline{Q}_{\alpha a},\overline{\lambda }_{\alpha
},\overline{u}_{\alpha },\ a=1,2\}.  \label{s2}
\end{equation}%
It is well known that the Lagrangian dynamics is generated in a
"anticanonical" structure. The generator of the Lagrangian BRST symmetry is%
\begin{equation*}
S=S_{0}[q,u]+\cdots 
\end{equation*}%
Because $S_{0}$ is unique one we will consider that $S$ is unique and we
will introduce two antibracket structures which have the same properties
like in the standard theory:%
\begin{equation*}
(F,G)_{a}=\frac{\delta ^{r}F}{\delta Q^{\mathcal{A}}}\frac{\delta ^{l}G}{%
\delta Q_{\mathcal{A}a}^{\ast }}-\frac{\delta ^{r}F}{\delta Q_{\mathcal{A}%
a}^{\ast }}\frac{\delta ^{l}G}{\delta Q^{\mathcal{A}}}.
\end{equation*}%
The functionals $F$ and $G$ are depend to $Q^{\mathcal{A}}\ $and $Q_{%
\mathcal{A}a}^{\ast }$. On the basis of the graduation properties and of the
Grassmann parities \cite{cls7'} we define the paires canonical conjugate in
respect with these antibrackets%
\begin{equation}
(Q_{\mathcal{A}b}^{\ast },Q^{\mathcal{B}})_{a}=-\delta _{\mathcal{A}}^{%
\mathcal{B}}\delta _{ba}  \label{s3}
\end{equation}%
where $Q_{\mathcal{A}a}^{\ast }$ are expressed by (\ref{s1}) and $Q^{%
\mathcal{A}}$ are the fields of the theory (real $q^{i},u^{\alpha }$ and
ghosts $Q^{\alpha a},\lambda ^{\alpha }$)%
\begin{equation}
Q^{\mathcal{A}}\equiv \{Q^{A},u^{\alpha }\}=\{q^{i},Q^{\alpha a},\lambda
^{\alpha },u^{\alpha },\ \alpha =1,2\}.  \label{s4}
\end{equation}%
For the Lagrange multiplyers we will have%
\begin{equation*}
\varepsilon (u^{\alpha })=\varepsilon _{\alpha },gh(u^{\alpha })=0.
\end{equation*}%
We observe that some antifields of the theory have a canonical conjugate in
the antibracket structure (\ref{s3}) and other antifields not have canonical
pair (\ref{s2}). So, the BRST differentials $\{s^{a},\ a=1,2\}$ will have
the following decomposition%
\begin{equation}
s^{a}\ast =(s^{a})^{can}\ast +V^{a}\ast =(\ast ,S)^{a}\ast +V^{a}\ast ,\
a=1,2  \label{s5}
\end{equation}%
where the non-canonical operators $V_{a}$ have the form \cite{cls7'}%
\begin{equation}
V^{a}\ast \equiv (-)^{\varepsilon (Q^{\mathcal{A}})}\varepsilon ^{ab}Q_{%
\mathcal{A}b}^{\ast }\frac{\delta ^{r}}{\delta \overline{Q}_{\mathcal{A}}}%
\ast .\text{ }  \label{an26}
\end{equation}%
The nilpotency condition for $s^{a}$ (\ref{s5}) leads to the master
equations (\ref{03}). For our irreducible theory, the proper solution of the
master eqs (\ref{03}) till terms linear in the antifields is 
\begin{equation*}
S=S_{0}+\int dt\ (q_{ia}^{\ast }a_{\alpha }^{i}Q^{\alpha a}+u_{\alpha
a}^{\ast }\overset{.}{Q}^{\alpha a}-u_{\alpha a}^{\ast }(V_{\beta }^{\alpha
}+G_{\delta }U_{\beta }^{\delta \alpha })Q^{\beta a}+
\end{equation*}%
\begin{equation*}
+u_{\alpha a}^{\ast }(f_{\beta \gamma }^{\alpha }+G_{\delta }g_{\beta \gamma
}^{\delta \alpha })u^{\gamma }Q^{\beta a}(-)^{\varepsilon _{\beta
}}+Q_{\alpha ab}^{\ast }\left( \varepsilon ^{ab}\lambda ^{\alpha }+\frac{1}{2%
}(-)^{\varepsilon _{\beta }}(f_{\beta \gamma }^{\alpha }+G_{\delta }g_{\beta
\gamma }^{\delta \alpha })Q^{\gamma b}Q^{\beta a}\right) +
\end{equation*}%
\newline
\begin{equation*}
+\lambda _{\alpha a}^{\ast }\frac{1}{2}(-)^{\varepsilon _{\beta
}+1}(f_{\beta \gamma }^{\alpha }+G_{\delta }g_{\beta \gamma }^{\delta \alpha
})\lambda ^{\beta b}Q^{\gamma a}+\overline{q}_{i}\left( a_{\alpha
}^{i}\lambda ^{\alpha }+\frac{1}{2}a_{\beta }^{j}\frac{\delta a_{\alpha }^{i}%
}{\delta q^{j}}Q^{\alpha c}Q^{\beta b}\varepsilon _{bc}\right) +
\end{equation*}%
\begin{equation*}
+\overline{q}_{i}\frac{1}{12}(-)^{\varepsilon _{\beta }}(f_{\beta \sigma
}^{\alpha }+G_{\delta }g_{\beta \sigma }^{\delta \alpha })(f_{\gamma \rho
}^{\sigma }+G_{\delta }g_{\gamma \rho }^{\delta \sigma })\varepsilon
_{cd}Q^{\rho a}Q^{\gamma c}Q^{\beta d}+
\end{equation*}%
\begin{equation*}
+\overline{u}_{\alpha }\overset{.}{\lambda }^{\alpha }+\overline{u}_{\alpha
}\left( -(V_{\beta }^{\alpha }+G_{\delta }U_{\beta }^{\delta \alpha
})+(-)^{\varepsilon _{\beta }}(f_{\beta \gamma }^{\alpha }+G_{\delta
}g_{\beta \gamma }^{\delta \alpha })u^{\gamma }\right) \lambda ^{\beta }+
\end{equation*}%
\begin{equation}
+\overline{Q}_{\alpha a}\left( -\right) ^{\varepsilon _{\beta }+1}(f_{\beta
\sigma }^{\alpha }+G_{\delta }g_{\beta \sigma }^{\delta \alpha })(\lambda
^{\sigma }Q^{\beta a}-\frac{1}{6}(f_{\gamma \rho }^{\sigma }+G_{\delta
}g_{\gamma \rho }^{\delta \sigma })\varepsilon _{bc}Q^{\rho a}Q^{\gamma
c}Q^{\beta b}).  \label{s7}
\end{equation}%
On the basis of the graduation rules and Grassmann parities \cite{cls7}, 
\cite{cls7'} we can identify the following variables%
\begin{equation}
P_{\alpha a}\equiv u_{\alpha a}^{\ast },\pi _{\alpha }\equiv \overline{u}%
_{\alpha }.  \label{s7.0}
\end{equation}%
These identifications will be very useful in the gauge fixing procedure \cite%
{AnnderPhys}. On the basis of the identifications (\ref{s7.0}), we will have 
\begin{equation}
Y=f^{\alpha }(q^{i})\overline{u}_{\alpha }.  \label{s7.2}
\end{equation}%
The following relations are valid 
\begin{equation}
q_{ia}^{\ast }=\frac{\delta ^{r}}{\delta q^{i}}(\frac{1}{2}\varepsilon
_{ab}V_{b}Y)=-\frac{\delta ^{r}f^{\alpha }(q^{i})}{\delta q^{i}}u_{\alpha
a}^{\ast },  \label{s29}
\end{equation}%
\begin{equation}
\overline{q}_{i}=\frac{\delta ^{r}Y}{\delta q^{i}}=-\frac{\delta
^{r}f^{\alpha }(q^{i})}{\delta q^{i}}\overline{u}_{\alpha }  \label{s29.0}
\end{equation}%
\begin{equation}
u^{\alpha }=-\frac{\delta ^{L}}{\delta u_{\alpha a}^{\ast }}(\frac{1}{2}%
\varepsilon _{ab}V_{b}Y)=f^{\alpha }(q^{i})  \label{s30}
\end{equation}%
the remaining antifields vanishing because of the choice (\ref{s7.2}) for $Y.
$ The gauge fixed action will be%
\begin{equation*}
\overline{S}_{1Y}=\overline{S}_{1}[Q^{A},u_{\alpha a}^{\ast },\overline{u}%
_{\alpha },u^{\alpha }=f^{\alpha }(q^{i}),q_{ia}^{\ast }=-\frac{\delta
^{r}f^{\alpha }(q^{i})}{\delta q^{i}}u_{\alpha a}^{\ast },
\end{equation*}%
\begin{equation}
\overline{q}_{i}=\frac{\delta ^{r}Y}{\delta q^{i}}=-\frac{\delta
^{r}f^{\alpha }(q^{i})}{\delta q^{i}}\overline{u}_{\alpha }]  \label{s34}
\end{equation}%
It leads to an effective action which is $s_{a}$-invariant and that can be
further used in the path integral. This path integral can be written as%
\begin{equation}
Z_{Y}^{L}=\int \mathcal{D}Q^{A}\mathcal{D}u_{\alpha a}^{\ast }\mathcal{D}%
\overline{u}_{\alpha }\exp (i\overline{S}_{1Y}).  \label{s35}
\end{equation}%
We will introduce the condensed notations%
\begin{equation}
\phi ^{A}\equiv \{Q^{A},u_{\alpha a}^{\ast },\overline{u}_{\alpha
}\}=\{q^{i},Q^{\alpha a},\lambda ^{\alpha },u_{\alpha a}^{\ast },\overline{u}%
_{\alpha }\}  \label{s36}
\end{equation}%
and we will consider the following BRST transformations%
\begin{equation}
\phi ^{A}\rightarrow \phi ^{A\prime }=\phi ^{A}-(s_{a}\phi ^{A})\mu
_{a}(-1)^{\varepsilon _{A}}.  \label{s37}
\end{equation}%
where $\mu _{a}$ are small fermionic constant parameters. The superjacobian
of this transformations can be aproximate with supertrace (because it
involves both fermionic and bosonic fields, the jacobian is a
superdeterminant) and the new integrating measure will be%
\begin{equation}
\mathcal{D}\phi ^{A\prime }=\left( 1-((-1)^{\varepsilon _{A}}\frac{\partial
^{r}}{\partial \phi ^{A}}\frac{\partial ^{r}S}{\partial \phi _{Aa}^{\ast }}%
+V^{a}S)\mu _{a}\right) \mathcal{D}\phi ^{A}=(1-(\Delta ^{a}S)\mu _{a})%
\mathcal{D}\phi ^{A}.  \label{s38}
\end{equation}%
In the previous relation, the operators $\Delta ^{a},a=1,2$ have the form (%
\ref{04}).

Exprimam $\Delta ^{a}S\ $using (\ref{s7}):%
\begin{equation*}
\Delta ^{a}S=\left( -\frac{\partial ^{r}a_{\beta }^{i}}{\partial q^{i}}+%
\frac{1}{2}[(-1)^{\varepsilon _{\beta }(\varepsilon _{\alpha
}+1)}-(-)^{\varepsilon _{\alpha }(\varepsilon _{\beta }+1)}](f_{\beta \alpha
}^{\alpha }+G_{\delta }g_{\beta \alpha }^{\delta \alpha })\right) Q^{\beta
a}+
\end{equation*}%
\begin{equation*}
\varepsilon ^{ab}u_{\alpha b}^{\ast }[\left( -\delta _{\beta }^{\alpha }%
\frac{d}{dt}-(V_{\beta }^{\alpha }+G_{\delta }U_{\beta }^{\delta \alpha
}+(-)^{\varepsilon _{i}}\frac{\partial ^{r}f^{\alpha }(q^{i})}{\partial q^{i}%
}a_{\beta }^{i})+(f_{\beta \gamma }^{\alpha }+G_{\delta }g_{\beta \gamma
}^{\delta \alpha })f^{\gamma }(q)\right) \lambda ^{\beta }
\end{equation*}%
\begin{equation*}
-(-)^{\varepsilon _{i}}\frac{\partial ^{r}f^{\alpha }(q^{i})}{\partial q^{i}}%
\frac{1}{2}a_{\beta }^{j}\frac{\partial a_{\alpha }^{i}}{\partial q^{j}}%
Q^{\alpha c}Q^{\beta b}\varepsilon _{bc}+\frac{1}{12}(-)^{\varepsilon
_{\beta }}(f_{\beta \sigma }^{\alpha }+G_{\delta }g_{\beta \sigma }^{\delta
\alpha })\cdot
\end{equation*}%
\begin{equation*}
\cdot (f_{\gamma \rho }^{\sigma }+G_{\delta }g_{\gamma \rho }^{\delta \sigma
})\varepsilon _{cd}Q^{\rho a}Q^{\gamma c}Q^{\beta d})]-(-)^{\varepsilon
_{i}}\varepsilon ^{ab}u_{\alpha b}^{\ast }\{\frac{\partial ^{r}}{\partial
q^{i}}(\frac{1}{2}\varepsilon _{ab}V_{b}\delta Y)[a_{\alpha }^{i}\lambda
^{\alpha }+
\end{equation*}%
\begin{equation*}
+\frac{1}{2}a_{\beta }^{j}\frac{\partial a_{\alpha }^{i}}{\partial q^{j}}%
Q^{\alpha c}Q^{\beta b}\varepsilon _{bc}+\frac{1}{12}(-)^{\varepsilon
_{\beta }}(f_{\beta \sigma }^{\alpha }+G_{\delta }g_{\beta \sigma }^{\delta
\alpha })(f_{\gamma \rho }^{\sigma }+G_{\delta }g_{\gamma \rho }^{\delta
\sigma })\varepsilon _{cd}Q^{\rho a}Q^{\gamma c}Q^{\beta d}]-
\end{equation*}%
\begin{equation*}
-(f_{\beta \gamma }^{\alpha }+G_{\delta }g_{\beta \gamma }^{\delta \alpha })%
\frac{\delta ^{r}}{\delta u_{\gamma c}^{\ast }}(\frac{1}{2}\varepsilon
_{cd}V_{d}\delta Y)\}.
\end{equation*}%
The functions $G_{\delta }$ which appear in the previous relations are the
constraints in which the real momenta was substitute (on the basis of their
equations of motions) in function of $q,\overset{.}{q}$. In the same way
appear the functions $a_{\alpha }^{i}$ from $\{q^{i},G_{\alpha }\}$.

The antifields was eliminated by gauge fixing procedure (\ref{s29})-(\ref%
{s30}) but considering the gauge fixing functional of the form $Y+\delta Y$.
Then, $\Delta ^{a}S=0$ if%
\begin{equation*}
-\frac{\partial ^{r}a_{\alpha }^{i}}{\partial q^{i}}+\frac{1}{2}%
((-1)^{\varepsilon _{\alpha }(\varepsilon _{\beta }+1)}-(-)^{\varepsilon
_{\beta }(\varepsilon _{\alpha }+1)})(f_{\alpha \beta }^{\beta }+G_{\delta
}g_{\alpha \beta }^{\delta \beta })=0
\end{equation*}%
and if $\frac{\partial ^{r}}{\partial q^{i}}(\frac{1}{2}\varepsilon
_{ab}V_{b}\delta Y)$ and $\frac{\delta ^{r}}{\delta u_{\gamma c}^{\ast }}(%
\frac{1}{2}\varepsilon _{cd}V_{d}\delta Y)$ anuleaza termenii din paranteza
(cancel the terms) ce apare pe langa $\varepsilon ^{ab}u_{\alpha b}^{\ast }$%
. With other words, the measure remain invariant ($\Delta ^{a}S=0$) to the
transformation (\ref{s37}) if $\delta Y$ induced by this transformation in
the gauge fixing functional indeplineste cerintele de mai sus. If $\Delta
^{a}S\neq 0$, the operators $\Delta ^{a},a=1,2$ determine the $sp(2)$- QME
in the $sp(2)$ symmetric formulation of gauge theories, where the action $%
S_{Y+\delta Y}$ passes to a new "quantum action", $W,$ containing the $\hbar
-$order corrections and satisfying the equations (\ref{06}).

\section{Conclusions}

The problem of the quantum extension of the BRST procedure has been
investigated, both in the BFV and in the BV formalisms. The main questions
arising in our study were: What is the form of the quantum master equation?
How is looking and what is the meaning of the quantum action? How is it
possible to extend the standard BRST\ quantization procedure in order to
obtain $sp(2)$ or $sp(3)$ quantum master equations?

We tried to offer comprehensible answers to all these questions and to
illustrate them by considering the case of nonlinear gauge theories which
mix bosonic and fermionic fields and whose gauge superalgebra is defined by
nonlinear relations of the form (\ref{d0}),(\ref{d1}). For such a special
system considered in \cite{8} we implemented the Hamiltonian $sp(2)$
formalism and we pointed out the condition for which the path integrals are
BRST invariant and how the gauge fixing function has to be chosen in order
to assure the invariance when it is not intrinsec for the model.


\begin{thebibliography}{99}
\bibitem{1} C. Becchi, A. Rouet, R. Stora, \textit{Phys. Lett. }\textbf{B 52}%
\textit{\ }(1974) 344; I.V. Tyutin, \textit{Gauge Invariance in Field Theory
and Statistical Mechanics, }Lebedev Preprint 39 (1975), unpublished

\bibitem{BV} I.A. Batalin, G. A. Vilkovisky, \textit{Phys. Lett. }\textbf{B
69}\textit{\ }(1977) 309

\bibitem{BFV} E.S. Fradkin, G. A. Vilkovisky, \textit{Phys. Lett. }\textbf{B
55}\textit{\ }(1975) 224

\bibitem{HenTeit} M. Henneaux, C. Teitelboim, \emph{Quantization of Gauge
Systems}, Princeton Univ. Press (1992)

\bibitem{3} A. B. Zamolodchikov, \textit{Theor. Math. Phys.} \textbf{65}
(1986) 1205

\bibitem{4} C. M. Hull, \textit{Nucl. Phys.} \textbf{B353} (1991) 707; K.
Schoutens, A. Servin, P. van Nieuwenhuizen, \textit{Phys. Lett.} \textbf{B255%
} (1991) 549

\bibitem{Lav} M. Asorey, P. M. Lavrov, O. V. Radchenko, A. Sugamoto, \textit{%
arXiv:0809.3322v2}

\bibitem{cls7} I.A. Batalin, P.\thinspace M. Lavrov, I.V. Tyutin, \textit{J.
Math. Phys.}\textbf{31}\textit{\ }(1990) 1487

\bibitem{cls7'} Ph. Gregoire, M. Henneaux, \textit{Phys. Lett. }\textbf{B 277%
}\textit{\ }(1992) 459

\bibitem{noi} R. Constantinescu, C. Ionescu, \textit{Int.J. Mod.Phys.A }%
\textbf{21} (2006) 1567

\bibitem{5} K. Bering, \textit{Mod.Phys.Lett.} \textbf{A11} (1996) 499-513

\bibitem{7} A. Dresse, M. Henneaux, \textit{J. Math. Phys.} \textbf{35}
(1994) 1334

\bibitem{AnnderPhys} R. Constantinescu, C. Ionescu, \textit{Ann. der Physik} 
\textbf{15} (2006) 169

\bibitem{8} I. Batalin, R. Marnelius, \textit{Phys.Lett.} \textbf{B441}
(1998) 243-249
\end{thebibliography}
\end{document}